\documentclass[]{iopart}
\usepackage{iopams}  
\usepackage{psfig}  
\usepackage{times}

\begin{document}

\comment[General Relativistic Rossby-Haurwitz waves of a
	differentially rotating shell]{General
	Relativistic Rossby-Haurwitz waves of a slowly
	and differentially rotating
	%incompressible 
	fluid shell}

\author{Marek A. Abramowicz$^{1,2}$, Luciano
Rezzolla$^{1,3}$ and \\ Shin'ichirou Yoshida$^{1}$}

\address{~}

\address{$^1$SISSA, International School for Advanced Studies,
        Via Beirut 2, 34014 Trieste, Italy}

\address{$^2$Department of Astrophysics, Chalmers University,
        41296 G{\"o}teborg, Sweden}

\address{$^3$INFN, Department of Physics, University of
        Trieste, Via Valerio 2, 34127 Trieste, Italy}

\begin{abstract}
We show that, at first order in the angular velocity, the
general relativistic description of Rossby-Haurwitz waves
(the analogues of $r$ waves on a thin shell) can be
obtained from the corresponding Newtonian one after a
coordinate transformation. As an application, we show
that the results recently obtained by Rezzolla and
Yoshida (2001) in the analysis of Newtonian
Rossby-Haurwitz waves of a slowly and differentially
rotating, fluid shell apply also in General Relativity,
at first order in the angular velocity.
\end{abstract}

% stellar rotation, stellar oscillation, neutron stars, relativistic stars

\section{Introduction}

	The investigation of $r$ modes in rotating stars
has been the subject of a widespread interest in the
recent past. Among the numerous aspects considered so far
(see Andersson and Kokkotas 2001, as well as Friedman and
Lockitch 2001 for recent reviews), a particularly
intriguing one has focussed on the effects induced by the
differential rotation of the background stellar model
(Spruit 1999, Rezzolla {\it et al} 2000, 2001a, 2001b)
and whether this could prevent the excitation of the $r$
modes (Karino {\it et al} 2001, Lindblom {\it et al}
2001). To alleviate the complications introduced by
differential rotation in the $r$-mode eigenvalue problem,
Rezzolla and Yoshida (2001) have recently investigated
the properties of the analogues of $r$ waves, the
Rossby-Haurwitz waves (Haurwitz, 1940), on a
differentially rotating, Newtonian thin shell of
incompressible fluid. In this framework, the eigenvalue
problem is much simpler to solve, but incorporates many
of the mathematical properties of the corresponding
eigenvalue problems for multidimensional Newtonian stars
or for slowly-rotating relativistic stars. The most
important of these properties is the possible existence
of a singular behaviour for the eigenvalue problem. The
purpose of this Comment is to extend, in a simplified
model, the results of Rezzolla and Yoshida (2001) to the
general relativistic case.

\section{A useful coordinate transformation: the uniform
rotation case}

	We consider here a slowly rotating body and make
our calculations including terms up to first order in the
angular velocity. To this order, the object remains
spherical and Rossby-Haurwitz waves propagate in the
spherical shell representing its surface. If the body is
uniformly rotating and a spherical coordinate system $(t,
r, \theta, \phi)$ is used, the spacetime metric is given
by (Hartle and Thorne 1968)
\begin{equation}
\label{metric_1}
ds^{2} = e^{2\nu(r)}dt^2 - R^2 d\theta^2 - 
	R^2 \sin^2 \theta \left[d\phi - 
	\omega(r) dt \right]^2 \ ,
\end{equation}
and on the surface $r = R$, assumes the form
\begin{equation}
\label{metric_1R}
ds^{2} = e^{2\nu_{_{\rm R}}}dt^2 - R^2 d\theta^2 - 
	R^2 \sin^2 \theta \left[d\phi - 
	\omega_{_{\rm R}}dt \right]^2 \ ,
\end{equation}
where, the redshift $\nu_{_{\rm R}}\equiv\nu(R)$ and the
frame dragging angular velocity $\omega_{_{\rm
R}}\equiv\omega(R)$ are constants. Consider now the
simple coordinate transformation
\begin{eqnarray}
\label{coord_trans_0}
t &\ \longrightarrow \ \ {\widetilde t} & = e^{\nu_{_{\rm R}}}t \ , \\
\label{coord_trans_1}
\phi &\ \longrightarrow \ \ {\widetilde \phi} & = \phi - \omega_{_{\rm R}} t \ , \\
\label{coord_trans_2}
\theta & \ \longrightarrow \ \ {\widetilde \theta} & = \theta\ ,
\end{eqnarray}
\noindent which brings the metric (\ref{metric_1R}) into
the evidently Minkowski form
\begin{equation}
ds^2 = d{\widetilde t^{\;2}} - R^2 ( d{\widetilde
       \theta}^2 + \sin^2 {\widetilde \theta}
       d{\widetilde \phi}^2)  \ .
\end{equation}
The coordinate transformation
(\ref{coord_trans_0})--(\ref{coord_trans_2}) shows that,
at first order in $\Omega$ and limited to the spherical
surface $r=R$, the general relativistic description of
the physics on the slowly and uniformly rotating thin
shell can be done entirely in Newtonian terms. We will
exploit this feature to deduce, from known Newtonian
results, the properties of general relativistic
Rossby-Haurwitz waves in a slowly, differentially
rotating fluid shell. Hereafter, we will indicate with a
tilde all of the Newtonian quantities.

	As a first application of this procedure, we
recall that for a shell rotating with a uniform angular
angular velocity ${\widetilde \Omega}$, the Newtonian
dispersion relation for the Rossby-Haurwitz modes at
first order in ${\widetilde \Omega}$ takes the form
(Haurwitz 1940; Stewartson and Rickard 1969)
\begin{equation}
\label{newt_disp_rel}
{\widetilde\Pi} = {\widetilde \Omega} - 
	\frac{2 {\widetilde \Omega}}{l (l + 1)} \ .
\end{equation}
Here ${\widetilde \Pi} \equiv {\widetilde \sigma}/m$ is
the phase velocity\footnote{In Rezzolla and Yoshida
(2001) the phase velocity was indicated as $\omega_{\rm
ph}$, but this symbol is not used here to avoid confusion
with the frame dragging angular velocity $\omega$.} and
${\widetilde \sigma}$ is the eigenfrequency of the mode
with wavenumbers $m$ and $l$. Using the coordinate
transformation
(\ref{coord_trans_0})--(\ref{coord_trans_2}) we can write
\begin{equation}
\label{o_pi_tilde}
\fl \qquad {\widetilde \Omega} \equiv \frac{d\widetilde \phi}{d
	{\widetilde t}} = e^{-\nu_{_{\rm R}}}(\Omega - \omega_{_{\rm R}}) \ ,
\qquad \qquad 
{\widetilde \Pi} = e^{-\nu_{_{\rm R}}}(\Pi- \omega_{_{\rm R}}) \ .
\end{equation}
Inserting now expressions (\ref{o_pi_tilde}) in
(\ref{newt_disp_rel}) we immediately obtain
\begin{equation}
\label{gr_disp_rel}
{\Pi} = \Omega - \frac{2(\Omega - \omega_{_{\rm R}})}
	{l (l + 1)} \ , 
\end{equation}
which represents the general relativistic expression for
the Newtonian dispersion relation
(\ref{newt_disp_rel}). As expected, our expression
(\ref{gr_disp_rel}) agrees with the dispersion relation,
evaluated on the surface $r=R$, and derived by Kojima
(1997, 1998; see also Beyer and Kokkotas 1999; Lockitch
{\it et al} 2001), through direct but lengthy
relativistic calculations.

\section{The differential rotation case}

	Although the rotating body generating the line
element (\ref{metric_1}) is uniformly rotating, we allow
for a differential rotation ${\widetilde \Omega} =
{\widetilde \Omega} (\mu)$, with $\mu \equiv \cos\theta$,
be present on the thin shell\footnote{If the underlying
body is also rotating differentially the frame dragging
angular velocity $\omega$ acquires a $\theta$-dependence
already at first order in the angular velocity. In this
case $\omega$ needs to be expressed in a series
expansion in terms of vector spherical harmonics and the
line element (\ref{metric_1}) to be suitably corrected
(Hartle 1970).}. Already for this simplified model, an
analytic dispersion relation for the Rossby-Haurwitz
waves propagating on the shell is not available, not even
in the Newtonian limit. However, we can express the
eigenvalue problem for the waves on a differentially
rotating thin shell of radius $r = R$ through the
ordinary differential equation (Rezzolla and Yoshida
2001),
\begin{eqnarray}
\label{eigen_eq}
\fl \frac{d}{d\mu}\left[(1-\mu^2)\frac{d{\widetilde \chi}}{d\mu}\right]
        - \frac{m^2}{1-\mu^2}{\widetilde \chi}
        - \frac{2{\widetilde \Omega}}
	{{\widetilde \Pi}-{\widetilde \Omega}}
        \left[1 + \left(\frac{2\mu}{{\widetilde \Omega}}\right)
	\frac{d{\widetilde \Omega}}{d\mu}
        - \left(\frac{1-\mu^2}{2{\widetilde \Omega}}\right)
        \frac{d^2{\widetilde \Omega}}{d\mu^2}\right]
	{\widetilde \chi} = 0\ , 
\end{eqnarray}
where ${\widetilde \chi}(\mu)/\sin\theta = u^{\theta}$ is
the eigenfunction of the mode, with $u^{\theta}$ being
the $\theta$ component of the velocity
perturbation. Because of the coordinate transformation
(\ref{coord_trans_0})--(\ref{coord_trans_2}), the
numerical solution of the general relativistic analogue
of equation (\ref{eigen_eq}) is not necessary. Rather,
once the (tilde) Newtonian solution of (\ref{eigen_eq})
is known, the properties of general relativistic
Rossby-Haurwitz waves of a slowly and differentially
rotating fluid shell at first order in the shell's
angular velocity are simply determined as
\begin{eqnarray}
\label{gr_res_0}
\Omega(\mu) &=& e^{\nu_{_{\rm R}}}{\widetilde \Omega}(\mu) 
	+ \omega_{_{\rm R}}\ , \\
\label{gr_res_1}
\sigma &=& e^{\nu_{_{\rm R}}}{\widetilde \sigma} 
	+ m \omega_{_{\rm R}}\ , \\
\label{gr_res_2}
\chi &=& {\widetilde \chi}\ .
\end{eqnarray}
An important issue concerning the eigenvalue problem for
Rossby-Haurwitz waves is that of ``corotation'', i.e. of
whether differential rotation could bring the ratio
${\widetilde \Pi}/{\widetilde \Omega}(\mu)$ to be one. If
this would happen, the denominator of the third term in
equation (\ref{eigen_eq}) would vanish making the
eigenvalue problem a singular one and preventing the
existence of the modes.

	In the case the shell is {\it uniformly rotating}
at an angular frequency $\Omega$, it is straightforward
to show that in General Relativity and at first order in
$\Omega$, corotation cannot take place since in this case
$\omega_{_{\rm R}}/ \Omega<1$ and expression
(\ref{gr_disp_rel}) gives
\begin{equation}
1 - \frac{\Pi}{\Omega} = \frac{2}{l(l+1)}
	\left( 1 - \frac{\omega_{_{\rm R}}}
	{\Omega}\right) > 0 \ .
\end{equation} 
	We next consider the shell to be {\it
differentially rotating} and whether corotation can occur
for a sufficiently large degree of differential
rotation. For a number of different laws of differential
rotation, Rezzolla and Yoshida (2001) have shown that the
solution of equation (\ref{eigen_eq}) does not show
evidence of corotation and that ${\widetilde \Pi}/
{\widetilde \Omega_{_{\rm E}}} <1$ even for
asymptotically large values of differential rotation,
with ${\widetilde \Omega_{_{\rm E}}} \equiv {\widetilde
\Omega}(\mu=0)$ being the minimum angular velocity. Using
expressions (\ref{o_pi_tilde}) it is straightforward to
deduce that this is true also for the general
relativistic analogue of equation (\ref{eigen_eq}), for
which
\begin{equation}
1 - \frac{\Pi}{\Omega_{_{\rm E}}} =
	\frac{e^{\nu_{_{\rm R}}}{\widetilde\Omega_{_{\rm E}}}}
	{e^{\nu_{_{\rm R}}}{\widetilde \Omega_{_{\rm E}}} 
	+ \omega_{_{\rm R}}}\left( 1 - \frac{\widetilde \Pi}
	{{\widetilde \Omega_{_{\rm E}}}}\right) > 0 \ .
\end{equation} 
As a result, the general relativistic eigenvalue problem
for Rossby-Haurwitz waves for the line element
(\ref{metric_1}) does not show any singular behaviour.

	While we regard this result as interesting and
indicative, it does not necessarily imply that the
eigenvalue problem for $r$ waves cannot be singular for a
uniformly, slowly rotating star (see Kojima 1998, but
also Ruoff and Kokkotas 2001a, 2001b, and Yoshida
2001). This is because, by construction, the general
relativistic equivalent of equation (\ref{eigen_eq})
cannot account for the radial dependence of the frame
dragging angular velocity $\omega$.  This is at the
origin of a singular behaviour reported at first order in
the slow-rotation approximation, even for uniformly
rotating stars, but which seems to disappear when terms
of ${\cal O}(\Omega^2)$ are taken into account (Yoshida
and Lee 2001). In other words, the singular eigenvalue
problem due to differential rotation on a shell and the
singular eigenvalue problem due, in the slow-rotation
approximation, to the frame dragging angular velocity,
are mathematically similar but physically distinct.

	As a final remark we underline that the procedure
followed here is generic and the coordinate
transformation
(\ref{coord_trans_0})--(\ref{coord_trans_2}) can be used
any time that first order general relativistic effects
need to be considered on a thin, slowly rotating shell.

\ack We would like to thank John Miller for useful
discussions and comments. This research has been
supported by the MIUR, by the EU Programme ``Improving
the Human Research Potential and the Socio-Economic
Knowledge Base" (Research Training Network Contract
HPRN-CT-2000-00137), and by the Swedish NFR grant.

\section*{References}
\begin{harvard}

\item[]Andersson N and Kokkotas K~D~2001 {\it
	Int. J. Mod. Phys. D} {\bf 10} 381 

\item[]Beyer H~R and Kokkotas K~D~1999 {\it
	Mon.~Not.~Roy.~Astron.~Soc.} {\bf 308} 745

\item[]Friedman J~L~ and Lockitch K~H~2001 {\it Proc. IX
	Marcel Grossman Meeting} ed V Gurzadyan, R Jantzen,
	R Ruffini (Singapore:  World Scientific)

\item[](-------- {\it Preprint} {\tt gr-qc/0102114})

\item[]Hartle J~B and Thorne K~S 1968
        {\it Astrophys. Journ.} {\bf 153} 807

\item[]Hartle J~B 1970
        {\it Astrophys. Journ.} {\bf 161} 111

\item[]Haurwitz B 1940 {\it J. Mar. Res.} {\bf 3}, 254

\item[]Karino K Yoshida S'i and Eriguchi Y 2001 \PR D
	{\bf 64} 024003

\item[]Kojima Y 1997 {\it Prog. Theor. Phys. Suppl.}
	{\bf 128} 251

\item[]Kojima Y 1998 {\it Mon. Not. Roy. Ast. Soc.}
	{\bf 293} 49

\item[]Lindblom L, Tohline J E and Vallisneri M~
	{\it Preprint} {\tt astro-ph/0109352}

\item[]Lockitch K H, Andersson N and Friedman J~L
	2001 \PR D {\bf 63} 024019

\item[]Rezzolla L, Lamb F~K~ and Shapiro
	S.L. 2000 {\it Astrophys. J.} {\bf 531} L139

\item[]Rezzolla L, Lamb F~K~, Markovic D and Shapiro
	S~L~ 2001a {\it Phys. Rev. D} {\bf 64} 104013

\item[]\dash 2001b {\it Phys. Rev. D} {\bf 64} 104014

\item[]Rezzolla L and Yoshida S'i 2001 \CQG
	{\bf 18} L87
 
\item[]Ruoff J and Kokkotas K~D 2001a {\it
	Mon. Not. Roy. Ast. Soc.} {\bf 328} 678

\item[]Ruoff J and Kokkotas K~D 2001b 
	{\it Preprint} {\tt gr-qc/0106073}

\item[]Spruit H~C 1999 {\it Astron. Astrophys.} {\bf 341} L1

\item[]Stewartson K and Rickard J A  1969 {\it
	J. Fluid. Mech.} {\bf 35} 759

\item[]Yoshida S 2001 {\it Astrophys. Journ.} {\bf 558} 263

\item[]Yoshida S and Lee U {\it Preprint} {\tt gr-qc/0110038}

\end{harvard}

\end{document}